\documentclass[10pt]{amsart}

\usepackage{bm}
\usepackage{graphicx}
\usepackage{float}
\usepackage{subfig}
\usepackage{cite}

\theoremstyle{plain}

\theoremstyle{definition}

\author{Joe Rusinko}
\address{Department of Mathematics, Winthrop University, Rock Hill SC
29733, USA}
\email{rusinkoj@winthrop.edu}
\author{Brian Hipp}
\address{Department of Mathematics, Winthrop University, Rock Hill SC
29733, USA}
\email{hippb@winthrop.edu}
\keywords{quartet puzzling, phylogenetics, invariants}
\title[Invariant Based QP]{Invariant Based Quartet Puzzling}
\date\today

\begin{document}

\begin{abstract} Traditional Quartet Puzzling algorithms use maximum likelihood methods to reconstruct quartet trees, and a puzzling algorithm to combine these quartets into a tree for the full collection of $n$ taxa.  We propose a variation of Quartet Puzzling in which the quartet trees are reconstructed using biologically symmetric invariants.   We find that under certain conditions, invariant based quartet puzzling outperforms Quartet Puzzling using maximum likelihood.  
\end{abstract}
\maketitle
\section{Introduction}
Invariant based reconstruction of phylogenetic trees was introduced by Cavender and Felsenstein \protect\cite{cavender}, Lake \protect\cite{lake} and Evans and Speed \protect\cite{evans}.   Invariants are relationships which proportions of observed data should satisfy if they evolved under a given tree and model of evolution. Invariant based reconstruction was originally found to be less effective at reconstruction than more traditional methods. (\protect\cite{huelsenbeck}, \protect\cite{jin}).  These shortcomings were to be expected given that initial efforts in this area did not use all possible expected relations among observed data.  Recent work by algebraic geometers has led to the construction of complete lists of phylogenetic invariants for certain models of evolution (\protect\cite{toricideal} , \protect\cite{invgenmark} ,\protect\cite{smalltrees}).  

Casanellas and Fernandez-Sanchez used a complete list of invariant equations to analyze the performance of invariant based reconstruction on quartets and found that they performed quite well, in some instances out performing traditional methods \protect\cite{quartetinv}.  The direct use of invariants on larger data sets would require the construction of a potentially enormous list of polynomial equations, which must then be evaluated on the possible evolutionary trees. The computational power required to complete these tasks makes the direct use of invariants impractical for large data sets, and thus some modifications must be introduced. We propose using a variation of quartet puzzling which uses invariants to compute the individual quartets, thus allowing the application of invariants to larger data sets.  

Strimmer and von Haeseler introduced quartet puzzling as a way to use the theoretical power of maximum likelihood while limiting the computational costs involved in a full maximum likelihood reconstruction \protect\cite{origqpuz}.  Subsequently, the use of quartet puzzling has become standard through programs such as TREE-PUZZLE \protect\cite{treepuzz}.  Quartet puzzling computes the optimal four taxa trees for every subset of four taxa from the data set.  A puzzling algorithm is used to combine these quartet trees into an overall tree for all of the taxa.  Simulation studies of quartet puzzling revealed that errors made in the choice of the individual quartets were propagated throughout the puzzling process, which in some instances caused inaccurate tree reconstructions \protect\cite{qdata}.  Attempts to reduce these errors include variations on quartet puzzling which limit the quartets that are examined, provide different weights for individual quartets, or modify the puzzling procedure (\protect\cite{shortqpuz},  \protect\cite{wo}, \protect\cite{qcleaning}).  Although it was conjectured that quartet based methods of reconstruction could not compete with neighbor joining algorithms for accuracy \protect\cite{wo}, the short quartet puzzling method of Snir (et. al.) disproved this claim by outperforming the neighbor joining algorithm \protect\cite{shortqpuz}.

In this paper we test a variation of quartet puzzling which is very similar to the original method proposed in \protect\cite{origqpuz}.  Instead of using maximum likelihood to reconstruct the quartet trees, we use invariant based reconstruction.  Our invariant reconstruction method follows \protect\cite{quartetinv} with the following modifications.  Instead of using the equations from the Algebraic Kimura model of evolution, we use a modification of the algebraic Jukes-Cantor equations for an unrooted quartet tree without molecular clock restrictions constructed in \protect\cite{toricideal} and available in \protect\cite{smalltrees}.  We made a modification to this list of equations by constructing a new set of equations we call \emph{biologically symmetric invariants} or BSI.  Following \protect\cite{quartetinv} we selected the quartet for which the average of the absolute values of the BSI was the lowest. 

As incorrect quartet propagation will still occur in the puzzling procedure, we would not expect this method to compete with variations on quartet puzzling.  Consequently we are testing whether invariant based methods perform comparably to maximum likelihood (ML) based quartet puzzling and would thus provide evidence that further investigation of invariant based reconstruction, either in full or through puzzling, might be fruitful. 

\section{Methods}
 Given a choice of model of evolution and a tree $T$, phylogenetic invariants are relationships satisfied by expected pattern frequencies in sequences evolving along $T$ under the given evolutionary model.  For our study we use the Jukes-Cantor Model of evolution and the unrooted quartet tree topology $(AB)(CD)$.  We denote the pattern frequencies  $p_{pattern}$ .  For example, in a particular four taxa data sample we use $p_{AAGC}$ to denote the frequency at which one observes the pattern $AAGC$ in the aligned nucleotide sequences.  Under the Jukes-Cantor model, the expected frequency of many of these observations are the same, which allows one to group pattern frequencies into classes $p_1,p_2,\cdots p_n$ for which the expected frequencies are the same (See \protect\cite{smalltrees} for explicit description).  Sturmfels and Sullivant found that the invariant equations are much easier to compute after a linear change of coordinates based on a discrete Fourier Transformation \protect\cite{toricideal}.  After the change of coordinates, they label the frequency classes $q_1, \cdots ,q_n$.

If the taxa have evolved under the Jukes-Cantor model along this tree, the expected relationships among the $q_i$ are called the model invariants.  One such invariant is $q_1*q_2=q_3*q_4$. Thus, if our data evolved under these hypotheses, we would expect $q_1*q_2-q_3*q_4$ to equal zero.  Casanellas and Fernandez-Sanchez used the sum of the absolute values of the evaluation of eact invariant equation as a score for how well a particular tree fits the data, and they selected the tree with the lowest score for phylogenetic reconstruction.  They found this method to be quite accurate in selecting the best quartet arrangement \protect\cite{quartetinv}.

The minimal collection of invariant equations used by Casanellas and Fernandez-Sanchez arises from a statistical framework, but because they were derived using algebraic geometry, this particular list has one drawback.  Biologically speaking we would like the trees $(AB)(CD)$ and $(BA)(CD)$ to have the same score since they represent the exact same relationship among the data.  This is not the case, however, because the invariant equations are non-symmetric.  The fact that two trees that represent the same relationship have different scores causes serious theoretical problems for phylogenetic inference, so we decided to address the asymmetric nature of the collection of invariants which produce these scores.  

To understand our solution to this problem we must first take a glimpse at the algebraic geometry involved in constructing the invariants.  Phylogenetic invariants are a collection of equations whose solution set describes a particular shape known as an algebraic variety.  There are many different collections of equations whose solution describes the same geometric shape. The invariants computed in \protect\cite{smalltrees}  are a minimal set of equations which define this variety.    

 To illustrate the theoretical difficulties involved with using this minimal set invariants we use an example with data drawn from the crab.meg file that comes with the software MEGA \protect\cite{mega}.  We computed the minimal invariant scores for each of the 24 orderings of the taxa A=Artemia\_salina ,B=Clibanarius\_vittatus, C=Paralithodes\_camtschatica and D=Pagurus\_acadianus.  Table \protect\ref{tradinvscores} lists selected scores from among these orderings. 
\begin{table}[h]
\begin{tabular}{|c|c|c|c|}
\hline
Ranking & Ordering & Score & Unrooted tree \\
\hline
1& DBCA & .00625529 & (AC)(BD) \\
2& CADB & .00634515 & (AC)(BD) \\
3& BCDA & .00638956 & (AD)(BC) \\
4& CBAD & .00640981 & (AD)(BC) \\
5& CBDA & .00641121 & (AD)(BC) \\
6& DABC & .00659321 & (AD)(BC) \\
7& ADCB & .00661345 & (AD)(BC) \\
8& DACB & .00661486 & (AD)(BC) \\
9& BDAC & .00673539 & (AC)(BD) \\
10& BCAD & .00675968 & (AD)(BC) \\
14 & BDCA & .00685342 &(AC)(BD)\\
15& ADBC & .00696333 & (AD)(BC) \\
16 & CABD & .00699116 & (AC)(BD) \\
19 & DBAC & .00719732 & (AC)(BD) \\
20 & ACDB & .00733505 & (AC)(BD)\\
\hline
\end{tabular}
\caption{Minimal invariant scores for selected orderings of a four species subset of crab data.}
\label{tradinvscores}
\end{table}
 Notice that while the best two overall scores correspond to the pairing $(AC)(BD)$, the following six best scores all point to $(AD)(BC)$ as the proper unrooted tree. To alleviate this problem we developed a set of invariant equations whose score corresponds only to the tree topology and does not depend on the underlying ordering of the data.  We do this in a manner which keeps the list of invariant equations as small as possible.

We used the following procedure to find a set of \emph{biologically symmetric invariants} or  \emph{BSI}.  We began with the list of invariants in $q$-coordinates for the Jukes-Cantor model of evolution on the unrooted tree $(AB)(CD)$ as found on the small trees website \protect\cite{smalltrees}.  For each of the biologically equivalent tree topologies (example $(BA)(CD)$) we identified  the corresponding change in $q$-coordinates.  We then made the change of coordinates in each of the invariant equations.  If any additional invariant equations appeared we added them to the list invariant equations.  After repeating this process for each equivalent tree, we ended up with 15 additional equations, bringing the total number of biologically symmetric invariant equations to 48.  When choosing a tree for a quartet $(A,B,C,D)$ we select the tree for which the average of the absolute value of all of the biologically symmetric invariants is the smallest.   When these new invariants are evaluated on the example data we get the scores in table \protect\ref{bsiscores}.  Using BSI  we would select the quartet tree  $(AC)(BD)$.
\begin{table}[h]
\begin{tabular}{|c|c|c|}
\hline
Ranking & Unrooted Tree & Score  \\
\hline
1 & (AC)(BD)& .005676\\
2 & (AD)(BC)& .005715 \\
3 & (AB)(CD)& .006237 \\
\hline
\end{tabular}
\caption{Biologically symmetric scores for a four species subset of crab data.}
\label{bsiscores}
\end{table}

\subsection{Invariant Based Quarter Puzzling Algorithm}
Our method follows the original quartet puzzling algorithm \protect\cite{origqpuzz}.  We begin by using biologically symmetric invariants to select the appropriate tree for each of the ${n\choose 4}$ quartets of data from our sample of $n$ taxa.  Due to the nature of the invariant scoring, it is highly unlikely that there will be a tie between scores.  Following the recommendations for the tree puzzling algorithm as described in the simulation results of \protect\cite{qdata}, we choose $1,000$ random orderings of the taxa.  For each ordering, we use BSI to select the best quartet tree for the first four taxa.  Additional taxa are added to the tree following the quartet puzzling algorithm \protect\cite{origqpuzz}.  When there is a tie among edges which are candidates for adding the additional taxa, an edge is selected at random from those tied as the most likely edge.  For each of the $1000$ orderings, the algorithm produces an unrooted bifurcating tree with $n$ taxa.  In the final step of reconstruction, we use the CONSENSE program, which is a part of the PHYLIP software package, to compute an unrooted consensus tree \protect\cite{phylip}.  Our reconstruction program is available upon request.

\subsection{Simulation Study}
Since quartet puzzling is the most similar model of phylogenetic reconstruction to our model, we tested our model using the data sets of Ranwez and Gascuel \protect\cite{qdata}. These data sets, which include 6 different tree topologies with 12 taxa, were used to analyze various quartet puzzling based algorithms and improvements. For each tree we have data sets of length $300$ and $600$ base pairs which were generated using the Seq-gen  software \protect\cite{seqgen} under the Kimura two-parameter model with a transition/transversion rate of 2.  Each tree and base pair length is run under four different assumptions of evolutionary rate.  The exact specifications appear in \protect\cite{qdata}, but can be described in terms of the average maximum pairwise distance (MD) .  The four data sets of low $MD \approx 0.1$,  medium $MD \approx 0.3$, fast $MD \approx 1.0$ and very fast $MD  \approx 2.0$ substitutions per site.

\begin{figure}
\includegraphics[width=.75\textwidth]{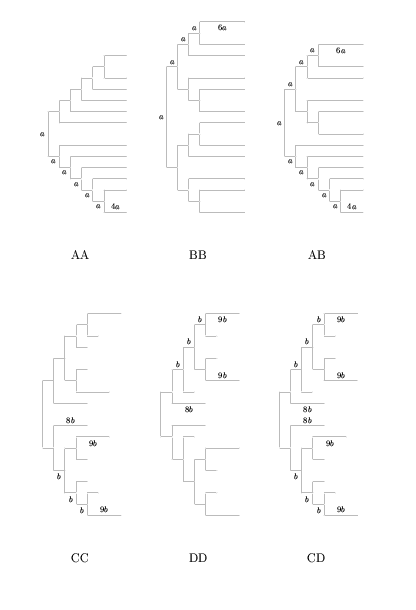}
\caption{Each interior branch is one unit long, $a$ for constant and $b$ for variable rate trees, and the lengths of external branches are given in multiples of $a$ and $b$.  Exact rates $a$ and $b$ are chosen in accordance to the evolutionary rate (see \protect\cite{qdata} for specific numbers).}
\label{trees}
\end{figure}

Three of the trees (AA, BB and AB) satisfied the molecular clock assumptions, while three trees CC, DD and CD did not.  Figure \protect\ref{trees}, taken from \protect\cite{qdata}, summarizes the trees. 

Following \protect\cite{qdata} we compared the results using the accuracy with which the algorithm reconstructed the exact tree, and the average Robinson-Foulds distance \protect\cite{robinson} between the reconstructed tree and the actual tree \protect\cite{robinson}.

\section{Results}
\subsection{Comparison with Maximum Likelihood Based Quartet Puzzling}
The results of our study are recorded in the four tables below. Table \protect\ref{l300acc} describes the accuracy of the algorithm in reconstructing the correct tree for length 300 sequences, while table \protect\ref{l300dist} shows the average Robinson-Foulds distance between the correct tree and the reconstructed tree.  Tables \protect\ref{l600acc}  and \protect\ref{l600dist} list the results for sequences of length 600 base pairs.   We list all data in relation to the results found on the same data set for the quartet puzzling algorithm,  listed as $Q_{4.2}$ in  \protect\cite{qdata}.
\begin{table}[h]
\begin{tabular}{|c|c|c|c|c|c|c|c|c|}
\hline
 &  AA & BB & AB & AVG & CC & DD & CD & AVG \\
\hline
M=0.1 BSI & 5 & 9 &8 & 7&11 & 10 &12  &   11\\

M=0.1 ML & 1 & 3 &2 &2 &3 & 3 & 3 & 4  \\
\hline
M=0.3 BSI & 13& 25 &17 &18 &33 &34  &  36&  34 \\

M=0.3 ML& 4 & 14 &7 &9 &18 & 24 & 21 & 21  \\
\hline
M=1.0 BSI &3 & 10 &2 &5 &20 &  23&21  & 21 \\

M=1.0 ML& 0& 3 &1 &2 &17 & 26 & 22 & 22  \\
\hline
M=2.0 BSI &0& 0 &0 &.0 &0 & 0 & 0 &  0 \\

M=2.0 ML &0& 0 &0 &.0 &1 & 3 & 1 & 2  \\
\hline
\end{tabular}
\caption{Length 300  BSI vs ML accuracy percentages}
\label{l300acc}
\end{table}

\begin{table}[h]
\begin{tabular}{|c|c|c|c|c|c|c|c|c|}
\hline
 &  AA & BB & AB & AVG & CC & DD & CD & AVG \\
\hline
M=0.1 BSI & 4.3 & 3.8 &4.1 &4.1 &4.5 & 4.5 &4.3  & 4.4  \\

M=0.1 ML & 4.3 & 3.8 &4.1 &4.1 &3.6 & 3.7 & 3.6 & 3.6  \\
\hline
M=0.3 BSI &3.3 & 2.7 &3.3 &3.1 & 2.1&2.1 &  2.1&  2.1 \\

M=0.3 ML &3.1 & 2.2 &2.9 &2.7 &1.8 & 1.7& 1.8 & 1.8  \\
\hline
M=1.0 BSI & 5.0 & 5.1 &5.6 &5.2 &3.1& 2.9 & 3.0 & 3.0  \\

M=1.0 ML & 4.3 & 3.6 &4.1 &4.0 &1.9 & 1.7 & 1.8 & 1.8  \\
\hline
M=2.0 BSI & 9.0& 11.7 &10.5 &10.4 &11.0 &11.5  & 11.1 &11.2   \\

M=2.0 ML & 6.6& 6.5 &6.6 &6.6 &4.3 & 4.1 & 4.2 & 4.2  \\
\hline
\end{tabular}
\caption{Length 300  BSI vs ML average Robinson-Foulds distance}
\label{l300dist}
\end{table}
\begin{table}[H]
\begin{tabular}{|c|c|c|c|c|c|c|c|c|}
\hline
 &  AA & BB & AB & AVG & CC & DD & CD & AVG \\
\hline
M=0.1 BSI&35&39 &35 & 36& 49& 45 &  49& 48 \\

M=0.1 ML &17& 29 &21 &22 &29 & 28 & 27 &28  \\
\hline
M=0.3 BSI & 55 & 65 &60 &60 &77 &79  & 79 & 78  \\

M=0.3 ML & 48 & 61 &54 &54 &74 & 79 & 78 & 77  \\
\hline
M=1.0 BSI &15 & 38 &22 &25 &67 &70 &66  &   68\\

M=1.0 ML& 21 & 43 &26 &30 &76 & 84& 79 & 80  \\
\hline
M=2.0 BSI&0 & 0 &1 &0 & 2&7 & 3 &  4 \\

M=2.0 ML& 0 & 1 &0 &1 &24 & 36 & 30 & 30  \\
\hline
\end{tabular}
\caption{Length 600  BSI vs ML accuracy percentages}
\label{l600acc}
\end{table}
\begin{table}[H]
\begin{tabular}{|c|c|c|c|c|c|c|c|c|}
\hline
 &  AA & BB & AB & AVG & CC & DD & CD & AVG \\
\hline
M=0.1 BSI & 1.9 & 1.8 &1.9 &1.9 & 1.4& 1.6 & 1.5 &  1.5 \\

M=0.1 ML& 1.9 & 1.5 &1.7 &1.7 &1.4 & 1.5 & 1.5 & 1.5  \\
\hline
M=0.3 BSI& 1.0 &0.9 &1.0 & 1.0& 0.5 & 0.5 & 0.5 &   0.5\\

M=0.3 ML& 0.8 & 0.6 &0.7 &0.7 &0.4 & 0.3 & 0.3 & 0.3  \\
\hline
M=1.0 BSI & 2.7 & 1.9 &2.6 &2.4 & 0.8&  0.8& 0.8 &  0.8 \\

M=1.0 ML& 1.8 & 1.0 &1.5 &1.4 &0.4 & 0.3 & 0.3 & 0.3  \\
\hline
M=2.0 BSI & 6.8 & 8.1 &7.7 &7.5 &7.3 &6.8  & 7.1 &   7.1\\

M=2.0 ML & 4.4 & 3.8 &4.3 &4.2 &1.7 & 1.4 & 1.6 & 1.6  \\
\hline
\end{tabular}
\caption{Length 600  BSI vs ML average Robinson-Foulds distance}
\label{l600dist}
\end{table}
\subsection{Speed Calculations}
To compare our method with traditional quartet puzzling, we ran the program with $N=1000$ random orderings of the data.  The number of orderings was chosen to match the conditions of the study in \protect\cite{qdata}.  We also tested the effect of using a smaller number of orderings on both the speed and reconstruction accuracy. The results are described in  Table \protect\ref{choiceofn}. We ran our program on a Dell Optiplex 960 which has an Intel 2 Duo 3 GHz processor and 3.5 GB of RAM.

\begin{table}[H]
\begin{tabular}{|c|c|c|c|c|c|}
\hline
 Tree & N=1000 & N=500 & N=100 & N=50 & N=10 \\
\hline
CC M=0.1 &11 &10 & 11&11 &10\\
\hline
CC M=0.3& 33& 33& 34& 32&27\\
\hline
CC M=1.0 &20 &19 &18&18 & 14\\
\hline
CC M=2.0 &0 &0 &0 &0&0 \\
\hline
CD M=0.1 & 12&13 &13 &12&11 \\
\hline
CD M=0.3& 36&33 &34 &33 &27\\
\hline
CD M=1.0 &21 &22 &20 &19& 14\\
\hline
CD M=2.0 &0 &0 &0 &0& 0\\
\hline
DD M=0.1 & 10& 10& 10&10& 9\\
\hline
DD M=0.3&34&34 &34 & 33&28\\
\hline
DD M=1.0 & 23&24&22 & 22& 18\\
\hline
DD M=2.0 & 0&0& 0& 0&0\\
\hline
Avg run time per tree(in seconds)& 11.3&5.8 &1.4 &0.8 &0.4 \\
\hline
\end{tabular}
\caption{Accuracy with N randomly generated orderings (length=300 bp)}
\label{choiceofn}
\end{table}
\subsection{Effect of Using Biologically Symmetric Invariants in Comparison to Minimal Invariants}
On these same data sets we compared the accuracy of quartet puzzling with BSI invariants to those using the minimal invariant equations.  Table \protect\ref{BSIvsTrad} lists the comparison of these methods both in terms of accuracy and speed.
\begin{table}[h]
\begin{tabular}{|c|c|c|}
\hline
 Tree & BSI & Min. Inv. \\
\hline
CC M=0.1 &11 &10 \\
\hline
CC M=0.3& 33&32\\
\hline
CC M=1.0 & 20&15 \\
\hline
CC M=2.0 &0 &0\\
\hline
CD M=0.1 &  12&10\\
\hline
CD M=0.3& 36&32\\
\hline
CD M=1.0 &21&17\\
\hline
CD M=2.0 &0&0\\
\hline
DD M=0.1 &10&9\\
\hline
DD M=0.3&34&32\\
\hline
DD M=1.0 &23&20\\
\hline
DD M=2.0 &0&0 \\
\hline
Avg run time per tree & 11.3 & 11.2\\
\hline
\end{tabular}
\caption{Accuracy BSI invariants vs minimal invariants N (length=300 bp)}
\label{BSIvsTrad}
\end{table}

\section{Conclusion}
Based on our simulated data sets BSI quartet puzzling more frequently reconstructs the correct evolutionary tree for trees constructed with low to medium evolutionary rates.  Even in instances where BSI outperforms ML in reconstruction accuracy, the average Robinson-Foulds distance \protect\cite{robinson} remains slightly larger.   It is unclear to us whether the failure to compete with ML puzzling for high rates of evolutionary chance is due to the use of invariants, or to our underlying use of the Jukes-Cantor model.  When the average minimum pairwise distance was around $2.0$ substitutions per site it seems more important to account for the differing  transition and transversion frequencies, as is allowed in the Kimura Models.  Thus, the causes of our method's poor reconstruction performance in this setting remains unclear.

The fact that BSI-puzzling outperforms ML-puzzling under any circumstances  is  somewhat surprising and encouraging, given that ML models are known to reconstruct the correct quartet trees with very high accuracy.  We see this as evidence that invariant based models of reconstruction may play an important role in practical phylogenetic reconstruction in addition to the role they currently play in helping to understand the theoretical possibilities of phylogenetic reconstruction problems \protect\cite{mixture}.

We believe variations on this algorithm along similar lines to those applied to ML-puzzling should dramatically improve performance.  The authors are currently in the process of implementing some of these improvements. 

The transition from invariants on the small trees website \protect\cite{smalltrees} to biologically symmetric invariants has a minor effect on improving reconstruction accuracy.  The extra time required to compute the more reasonable biologically symmetric invariants is made up for by slightly improved accuracy and dramatically improved peace of mind.

All comparisons with ML based quartet puzzling were made using consensus trees based on 1000 random taxa orderings.  This number was selected to match the number of runs used in the puzzling in the simulation study under comparison.  Our testing indicates that dramatic increases in speed could be gained with little to no effect on accuracy by running only 50 data orderings.  We assume that the 1000 ordering suggestion in the TREE puzzle guidelines is there for working with larger data sets and that similar savings in time would occur when running traditional puzzling for fewer orderings on this data set.  Given the tremendous increase in speed further investigation into the appropriate number of orderings which maximize a combination of speed and accuracy may be of merit.

\section{Acknowledgments}
We would like to thank Danielle Couture and Jessica Turek whose work in the Summer Research Experience in Mathematics at Winthrop University motivated the authors to pursue this study.

\bibliography{qpuz}{}

\begin{thebibliography}{10}

\bibitem{mixture}
S.~Rhodes~J. Allman, E.~Petrovic and S.~Sullivant.
\newblock Identifiability of two-tree mixtures for group-based models.

\bibitem{qcleaning}
T.~Kearney P. et~al. Berry~V., Jiang.
\newblock Quartet cleaning: improved algorithms and simulations.

\bibitem{smalltrees}
Marta Casanellas, Luis~David Garcia, and Seth Sullivant.
\newblock Catalog of small trees.
\newblock In {\em Algebraic statistics for computational biology}, pages
  291--304. Cambridge Univ. Press, New York, 2005.

\bibitem{invgenmark}
Allman E. and J.~Rhodes.
\newblock Phylogenetic invariants for the general markov model of sequence
  mutation.

\bibitem{phylip}
J.~Felsenstein.
\newblock Phylip (phylogeney inference package) version 3.6.

\bibitem{huelsenbeck}
J.~Huelsenbeck.
\newblock Performance of phylogenetic methods in simulations.

\bibitem{cavender}
Cavender J. and Felsenstein J.
\newblock Invariants of phylogenies in a simple case with discrete states.

\bibitem{jin}
Jin L. and Nei M.
\newblock Limitations of the evolutionary parsimony method of phylogenetic
  analysis.

\bibitem{lake}
J.~Lake.
\newblock A rate-independent technique for analysis of nucleic acid sequences:
  evolutionary parsimony.

\bibitem{quartetinv}
Casanellas M. and F.~Fernandez-Sanche.
\newblock Performance of a new invariants method on homogeneuous and
  nonhomogeneous quartet trees.

\bibitem{seqgen}
A.~Rambaut and N.. Grassly.
\newblock Seq-gen: an application for the monte carlo simulation of dna
  sequence evolution along phylogenetic trees.

\bibitem{qdata}
V.~Ranwez and O.~Gascuel.
\newblock Quartet-based phylogenetic inference: improvements and limits.

\bibitem{robinson}
D.F. Robinson and L.~R. Foulds.
\newblock Comparison of phylogenetic trees.

\bibitem{evans}
Evans S. and Speed T.
\newblock Invariants of some probability models used in phylogenetic infrences.

\bibitem{treepuzz}
Strimmer K. Vingron~M. Schmidt, H.A. and A.~von Haesler.
\newblock Tree-puzzle: maximum likelihood phylogenetic analysis using quartet
  and parallell computiong.

\bibitem{shortqpuz}
Sagi Snir, Tandy Warnow, and Satish Rao.
\newblock Short quartet puzzling: a new quartet-based phylogeny reconstruction
  algorithm.
\newblock {\em J. Comput. Biol.}, 15(1):91--103, 2008.

\bibitem{origqpuz}
K.~Strimmer and A.~von Haeseler.
\newblock Quartet puzzling: A quartet maximum-likelihood method of
  reconstructing tree topologies.

\bibitem{origqpuzz}
K.~Strimmer and A.~von Haesler.
\newblock Quartet puzzling: a quartet maximum-likelihood method for
  reconstructing tree topologies.

\bibitem{toricideal}
S.~Sturmfels and S.~Sullivant.
\newblock Toric ideals of phylogenetic invariants.

\bibitem{mega}
Peterson N. Stecher G. Nei~M. Tamura, K. and S.~Kumar.
\newblock Mega5: Molecular evolution genetics analysis using maximum
  likelihood, evolutionary distance, and maximum parsimony methods.

\bibitem{wo}
Ranwez V. and O.~Gascuel.
\newblock Quartet-based phylogenetic inference: Improvements and limits.

\end{thebibliography}
\bibliographystyle{plain}

\end{document}